# Spectroscopic Needs for Training of LSST Photometric Redshifts

**A white paper submitted to the National Research Council**
**Committee on a Strategy to Optimize the U.S. O/IR System in the Era of LSST**

Alexandra Abate[1], Jeffrey A. Newman[2], Samuel J. Schmidt[3], Filipe B. Abdalla[4], Sahar Allam[5], Steven W. Allen[6,7], Réza Ansari[8], Stephen Bailey[9], Wayne A. Barkhouse[10], Timothy C. Beers[11], Michael R. Blanton[12], Mark Brodwin[13], Joel R. Brownstein[14], Robert J. Brunner[15], Matias Carrasco Kind[15], Jorge L. Cervantes-Cota[16], Elliott Cheu[1], Nora Elisa Chisari[17], Matthew Colless[18], Johan Comparat[19], Jean Coupon[20], Carlos E. Cunha[21], Axel de la Macorra[22], Ian P. Dell'Antonio[23], Brenda L. Frye[24], Eric J. Gawiser[25], Neil Gehrels[26], Kevin Grady[26], Alex

---

[1] Physics Department, University of Arizona, 1118 East 4th Street, Tucson, AZ 85721, USA
[2] Department of Physics and Astronomy and PITT PACC, University of Pittsburgh, 3941 O'Hara St., Pittsburgh, PA 15260
[3] Dept. of Physics, University of California, One Shields Ave., Davis, CA 95616, USA
[4] Astrophysics Group, Department of Physics & Astronomy, University College London, Gower Place, London WC1E 6BT, UK
[5] Fermi National Accelerator Laboratory, MS 127, PO Box 500, Batavia, Illinois 60510, USA
[6] Dept. of Physics, Stanford University, 382 Via Pueblo Mall, Stanford, CA 94305, USA
[7] SLAC National Accelerator Laboratory, 2575 Sand Hill Road MS 29, Menlo Park, CA 94025, USA
[8] Université Paris-Sud, LAL-IN2P3/CNRS, BP 34 , 91898 Orsay Cedex, France
[9] Lawrence Berkeley National Laboratory, 1 Cyclotron Rd, Berkeley, CA 94720, USA
[10] Department of Physics and Astrophysics, University of North Dakota, Grand Forks, ND 58202, USA
[11] National Optical Astronomy Observatories, 50 N. Cherry Avenue, P.O. Box 26732, Tucson, AZ 85726, USA
[12] Department of Physics, New York University, 4 Washington Place, Room 424, New York, NY 10003, USA
[13] Department of Physics and Astronomy, University of Missouri at Kansas City, 5110 Rockhill Road, Kansas City, MO 64110, USA
[14] Department of Physics, The University of Utah, 115 S 1400 E, Salt Lake City, UT 84112, USA
[15] Dept. of Astronomy, University of Illinois, 226 Astronomy Building, MC-221, 1002 W. Green St., Urbana, IL 61801, USA
[16] Instituto Nacional de Investigaciones Nucleares (ININ), Apartado Postal 18-1027 Col. Escandón, México DF 11801, México
[17] Department of Astrophysical Sciences, Princeton University, 4 Ivy Lane, Peyton Hall, Princeton, NJ 08544, USA
[18] Research School of Astronomy and Astrophysics, The Australian National University, Canberra, ACT 2611, Australia
[19] Campus of International Excellence UAM+CSIC, Cantoblanco, E-28049 Madrid, Spain
[20] Astronomical Observatory of the University of Geneva, ch. d'Ecogia 16, 1290 Versoix, Switzerland
[21] Kavli Institute for Particle Astrophysics and Cosmology, 452 Lomita Mall, Stanford University, Stanford, CA 94305, USA
[22] Depto. de Fisica Teorica and Instituto Avanzado de Cosmologia (IAC), UNAM, Mexico City, Mexico
[23] Department of Physics, Brown University, Box 1843, 182 Hope Street, Providence, RI 02912, USA
[24] Department of Astronomy and Steward Observatory, University of Arizona, 933 North Cherry Avenue, Tucson, AZ 85721, USA
[25] Dept. of Physics & Astronomy, Rutgers, The State University of New Jersey, 136 Frelinghuysen Rd., Piscataway, NJ 08854, USA
[26] NASA Goddard Space Flight Center, Greenbelt, MD 2077, USA

Hagen[27], P. Hall[28], Andrew P. Hearin[29], Hendrik Hildebrandt[30], Christopher M. Hirata[31], Shirley Ho[32], Klaus Honscheid[33], Dragan Huterer[34], Željko Ivezić[35], Jean-Paul Kneib[36,37], Jeffrey W. Kruk[26], Ofer Lahav[4], Rachel Mandelbaum[32], Jennifer L. Marshall[38], Daniel J. Matthews[2], B. Ménard[39], Ramon Miquel[40], Marc Moniez[8], H. W. Moos[39], John Moustakas[41], Adam D. Myers[42], C. Papovich[38], John A. Peacock[43], Changbom Park[44], M. Rahman[39], Jason Rhodes[45], Jean-stephane Ricol[46], Iftach Sadeh[4], Anže Slozar[47], Daniel K. Stern[45], J. Anthony Tyson[3], A. von der Linden[6], R. Wechsler[6,7], W. M. Wood-Vasey[2], Andrew R. Zentner[2]

E-mail Addresses: abate@email.arizona.edu; janewman@pitt.edu (corresponding author); sschmidt@physics.ucdavis.edu

**Context:** This white paper is a summary of the photo-z training needs described in the Snowmass White Paper *Spectroscopic Needs for Imaging Dark Energy Experiments* by Newman et al., available at http://arxiv.org/abs/1309.5384. That white paper focuses on estimating the amount of spectroscopic redshift data required to enable photometric redshift (photo-z) measurements with future imaging dark energy surveys. It divides the applications of spectroscopy into *training*, i.e., the use of spectroscopic redshifts to improve algorithms and reduce photo-z errors; and *calibration*, i.e., the accurate characterization of biases and uncertainties in photo-z's, which is critical for dark energy inference. We summarize here the conclusions from that white paper relevant to the training of LSST photometric redshifts; a separate white paper (Schmidt et al.) focuses on calibration needs. We refer the reader to the Snowmass white paper for all references.

---

[27] Department of Astronomy & Astrophysics, The Pennsylvania State University, 525 Davey Lab, University Park, PA 16802, USA
[28] Department of Physics and Astronomy, York University, 4700 Keele Street, Toronto, ON, Canada
[29] Yale Center for Astronomy & Astrophysics, Yale University, New Haven, CT
[30] Argelander-Institut für Astronomie, Auf dem Hügel 71, 53121 Bonn, Germany
[31] Department of Astronomy, The Ohio State University, 140 West 18th Avenue, Columbus, OH 43210, USA
[32] McWilliams Center for Cosmology, Carnegie Mellon University, 5000 Forbes Avenue, Pittsburgh, PA 15213, USA
[33] Department of Physics, The Ohio State University, 140 West 18th Avenue, Columbus, OH 43210, USA
[34] Department of Physics, University of Michigan, 450 Church St., Ann Arbor, MI 48109, USA
[35] Astronomy Department, University of Washington, PAB 357, 3910 15th Ave NE, Seattle, WA, USA
[36] Laboratoire d'Astrophysique, Ecole Polytechnique Fédérale de Lausanne (EPFL), Observatoire de Sauverny, CH-1290 Versoix, Switzerland
[37] Laboratoire d'Astrophysique de Marseille - LAM, Université d'Aix-Marseille & CNRS, UMR7326, 38 rue F. Joliot-Curie, 13388 Marseille Cedex, France
[38] Department of Physics & Astronomy, Texas A&M University, College Station, TX 77843, USA
[39] Department of Physics and Astronomy, Johns Hopkins University, Baltimore, MD 21218, USA
[40] Institut de Fisica d'Altes Energies (IFAE), Edifici Cn, Universitat Autonoma de Barcelona, E-08193 Bellaterra (Barcelona) Spain
[41] Department of Physics and Astronomy, Siena College, 515 Loudon Road, Loudonville, NY 12211, USA
[42] Department of Physics & Astronomy, University of Wyoming, 1000 E. University, Dept. 3905, Laramie, WY 82071, USA
[43] Institute for Astronomy, University of Edinburgh, Royal Observatory, Edinburgh EH9 3HJ, UK
[44] School of Physics, Korea Institute for Advanced Study, 85 Hoegiro, Dongdaemun-gu, Seoul 130-722, Korea
[45] Jet Propulsion Laboratory, California Institute of Technology, 4800 Oak Grove Drive, Pasadena, CA 91109, USA
[46] Laboratoire de Physique Subatomique et de Cosmologie Grenoble, 53 rue des Martyrs, 38026 Grenoble Cedex, France
[47] Brookhaven National Laboratory, P.O. Box 5000, Upton, NY 11973-5000

*Training* of photometric redshifts constitutes the use of samples with known *z* to develop or refine algorithms, and hence to reduce the random errors on individual photometric redshift estimates. *The larger and more complete the training set is, the smaller the RMS error in photo-z estimates should be, increasing LSST's constraining power.* For instance, LSST delivers photo-z's with an RMS error of $\sigma_z = 0.02(1+z)$ for *i*<25.3 galaxies in simulations with perfect template knowledge and realistic photometric errors, whereas photo-z's in actual samples of similar S/N have delivered $\sigma_z \sim 0.05(1+z)$. With a perfect training set of galaxy redshifts, we could achieve the system-limited performance; this would improve the DETF figure of merit from LSST lensing+BAO studies by ~25%, with much larger impact on some extragalactic science (e.g., the identification of galaxy clusters). ***Better photometric redshift training will improve almost all LSST extragalactic science, and hence address a wide variety of decadal science goals.***

**Basic Requirements:** A minimum of 30,000 spectra (ideally ~$10^5$) spanning the full range of properties of LSST samples are required to accurately characterize objects in both the core and outlier regions of the photo-z error distribution. To mitigate the effects of sample/cosmic variance, these observations must span a <u>minimum</u> of 15 widely-separated fields that are at least ~20 arcminutes in diameter.

The exposure times required with existing telescopes are long. To obtain projections, we assume that the secure redshift success rate is a function only of spectral signal-to-noise ratio, and extrapolate from the DEEP2 Galaxy Redshift Survey. We find that ~100 hours of on-sky time with the DEIMOS spectrograph on Keck would be needed to obtain redshifts for ~75% of targets on a mask down to the *i*=25.3 magnitude limit of LSST weak lensing samples, while achieving ~90% secure redshift rates would require ~600 hours before overheads. In order to enable optimal LSST photo-z training in months rather than years, wider fields (as compared to DEIMOS), high multiplexing, and large telescope apertures are necessary.

We note that while these secure redshift rates may sound discouragingly low, statistically complete spectroscopy and perfect training are not absolutely needed, as any biases due to incomplete spec-z samples can be characterized via cross-correlation methods (see the companion white paper by Schmidt et al.). However, all new spec-z's provide additional information about galaxy SEDs and colors, improving algorithms and reducing the scatter in photo-z estimates about their true values. Our primary goal is to maximize the impact of the set of objects that may be used for training photo-z's given the constraints of scarce telescope time.

Given these requirements, we now respond to specific questions from the committee:

**Q1. What O/IR capabilities are you using, are you planning to use, and will you need through the LSST era?** The required survey time would be minimized by utilizing a multi-object spectrograph with a field of view at least 20 arcmin in diameter and a multiplex factor of at least 2000 on as large a telescope as possible. In order to maintain a high rate of secure redshift identifications at $z>1$, the spectrograph must have sufficient resolution to split the [OII] 3727 Å doublet, requiring spectral resolution ($R=\lambda/\Delta\lambda$) ≳ 4000. Ideally, it should provide wavelength coverage and good sky subtraction or OH suppression well into the near-infrared in order to maximize the fraction of objects with secure redshift measurements; lacking those capabilities, infrared spectroscopy from space would be highly valuable.

The most efficient US-based options for this would be the GMACS spectrograph with the MANIFEST fiber feed for the Giant Magellan Telescope or WFOS on the Thirty Meter Telescope. GMACS/MANIFEST could collect a 75% (90%) complete training sample down to $i$=25.3 (the LSST lensing limit) in as little as 5 months (2.6 years) depending upon final design; however, the fiber feed and wide-field capabilities that allow this efficiency will not be available at first light. The lower field of view and multiplexing of TMT/WFOS are limiting factors; it could collect a 75% (90%) complete training sample for LSST in around 2 years (11 years). For comparison, Subaru/PFS could obtain a 75% (90%) complete sample in roughly 1 year (7 years); *given that smaller fields of view than PFS are needed for this work, it is likely that a survey speed similar to this could be obtained with new instrumentation on Gemini*.

**Q2. Do you have access to the O/IR capabilities you currently need to conduct your research?** None of the currently-planned options for LSST training surveys are within the federal O/IR system. Hence, our access to this spectroscopy is a major open question. We expect that it could most efficiently be obtained by partnering with non-federal facilities to conduct surveys which simultaneously will enable photo-$z$ training and a wide variety of spectroscopic galaxy evolution science. A well-designed survey could do both and be of common interest for both LSST's and these telescopes' user communities. We note that *the importance of funding for LSST training surveys was highlighted in the US High Energy Physics community's Snowmass report* (http://arxiv.org/abs/1401.6085) *and reiterated in the follow-up P5 report*; as a result there may be non-NSF resources that could be leveraged.

**Q3. Comment on the need for the U.S. community's access to non- federal O/IR facilities up to 30 meters in size:** Almost all options for LSST training surveys would rely on telescopes outside the federal O/IR system. Hence, this access is critical.

**Q10. What types of scientific and observing coordination among the various NSF telescopes (including Gemini and LSST) and non-federal facilities are the most important for making scientific progress in the next 10-15 years? How can such coordination best be facilitated?** Photometric redshift training is generally of limited interest to TACs, but it impacts a huge variety of extragalactic science. In order to ensure that the spectroscopy needed for LSST science will be obtained, it will be important for synergies between photo-z training and other planned surveys to be identified and leveraged. Assembling a sufficient dataset without impacting any single facility too severely will require communication and coordination of plans amongst all the large surveys to be conducted with each next-generation facility, both internally and between facilities.

Time on the largest telescopes will be a precious and highly sought-after resource. It is in the interest of the community that maximum use be made of this time; options for parallel observations or the automated addition of targets to utilize unused fibers in users' configurations are worth considering, and could benefit this work, as LSST photo-$z$ training targets would be available over a full 20,000 square degrees of sky.